# USING MOBILE AGENTS FOR INFORMATION RETRIEVAL IN B2B SYSTEMS

**Felicia GÎZĂ[1], Cristina TURCU[2], Ovidiu SCHIPOR[3]**

*[1]felicia@eed.usv.ro, [2]cristina@eed.usv.ro, [3]schipor@eed.usv.ro*

***Abstract*** *This paper presents an architecture of an information retrieval system that use the advantages offered by mobile agents to collect information from different sources and bring the result to the calling user. Mobile agent technology will be used for determine the traceability of a product and also for searching information about a specific entity.*

## Introducere

Accesul la informaţie nu mai reprezintă o problemă în ultimii ani, când la dispoziţia utilizatorului există multe surse de informare. Totuşi, de multe ori, informaţia dorită se găseşte printre alte informaţii mai puţin importante sau chiar irelevante. În aceste condiţii, metodele clasice de căutare a informaţiilor nu mai sunt eficiente. Operaţiile de trimitere a cererilor către baze de date sau de căutare repetată prin paginile web au devenit foarte complexe şi mari consumatoare de timp.
Principalul scop al sistemelor de colectare a informaţiilor este de a pune la dispoziţia utilizatorului doar informaţiile dorite. La proiectarea acestor sisteme trebuie luată în considerare diminuarea costurilor implicate de procesul de preluare a tuturor datelor (căutarea, filtrarea, prezentarea). În 2001, Tomasz Orzechowschi propunea următoarea clasificare din punct de vedere structural a acestor sisteme [1]:

1. un tip de sistem în care informaţiile sunt localizate în baze de date specifice şi au caracter asemănător (figura 1). În această categorie pot fi menţionate sistemele cu bazele de date ale centrelor de cercetare (articole ştiinţifice, rezultate ale cercetărilor în curs de desfăşurare), bazele de date bancare (informaţii despre conturile utilizatorilor), bazele de date B2B (in care se găsesc informaţii despre produsele sau serviciile comercializate);
2. al doilea tip de sistem nu prezintă nici un fel de limitări pentru localizarea informaţiilor cerute. Utilizatorul poate să solicite căutarea informaţiilor în diverse locaţii (atât baze de date, cât şi pagini web).

## Prezentare sistem

Se consideră sistemul B2B_RFID [2] pentru managementul şi urmărirea trasabilităţii materialelor, subansamblelor şi produselor finite, pe întreg lanţul de aprovizionare şi desfacere.
*Sistemul B2B_RFID* face parte din prima categorie de sisteme datorită structurii şi caracteristicilor sale:

- fiecare firmă are un server de baze de date distribuite;
- structura bazei de date este generală (poate fi utilizată pentru orice tip de afacere);
- la nivelul unei firme există o reţea Intranet prin intermediul căreia utilizatorii locali accesează resursele;
- comunicaţia între firme este realizată prin intermediul Internet-ului.

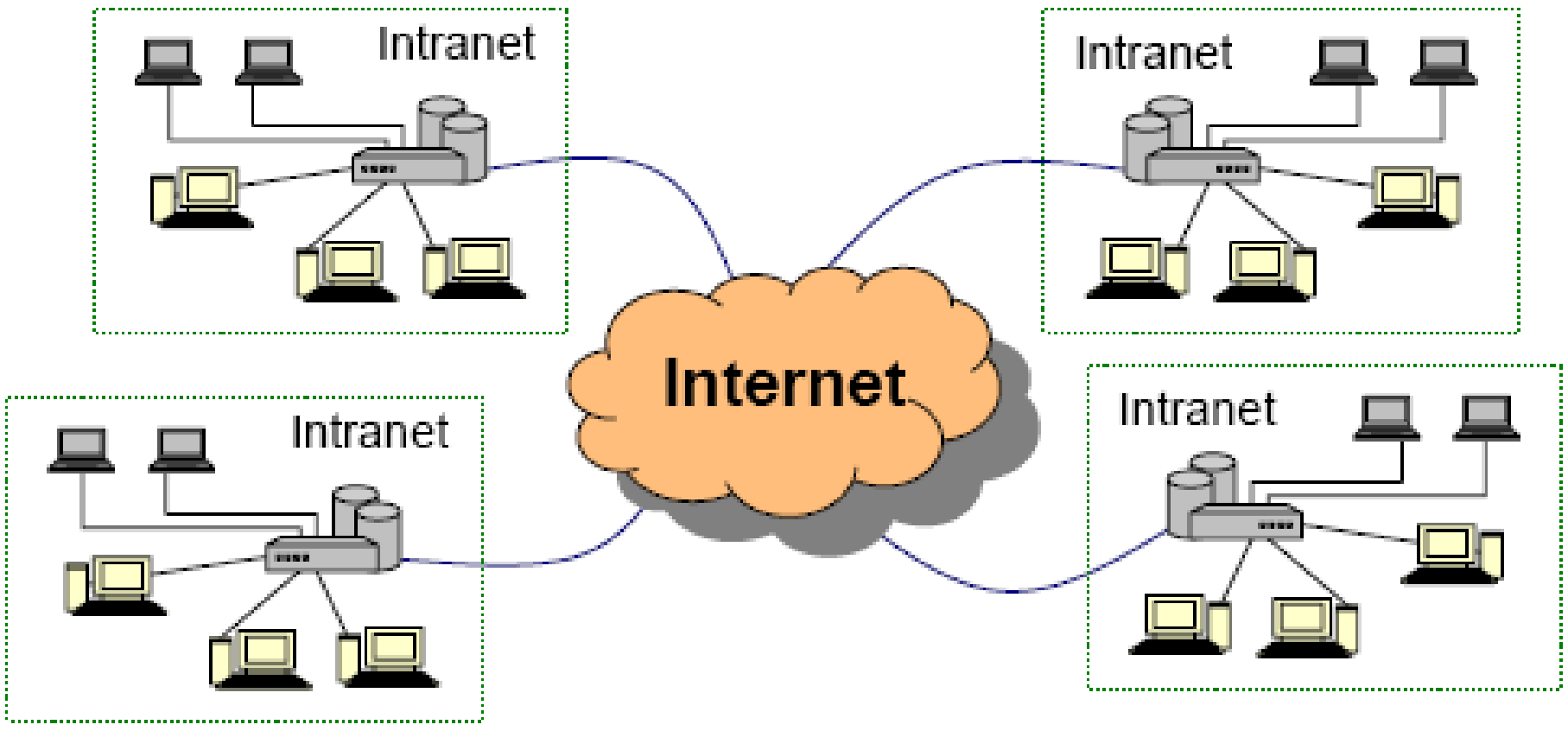

**Figura 1. Structura detaliata a primului tip de sistem de colectare a informaţiilor [1]**

Se observă că în cadrul acestor sisteme se disting două tipuri de comunicaţii:
- *interne* (în cadrul fiecărei reţele Intranet);
- *externe* (între reţele Intranet, prin intermediul Internet-ului)

*Comunicaţiile interne* se referă la comunicaţiile din cadrul fiecărei reţele Intranet de la nivelul firmei. În acest caz, modelul de bază este modelul client/server, în care persoana care caută informaţia este identificată ca şi client, iar serverul stochează bazele de date.
Alegerea unei tehnologii specifice de comunicaţie nu este cea mai importantă problemă în acest caz. Accesul la reţeaua locală nu necesită resurse financiare, iar lărgimea de bandă pentru reţelele locale este de obicei mare şi fără restricţii.
Principalul scop al *comunicaţiilor externe* este de a pune la dispoziţia partenerilor de afaceri şi filialelor aceleiaşi instituţii, informaţiile locale despre produsele şi serviciile proprii şi, implicit, de a avea acces şi la informaţiile despre produsele şi serviciile oferite de aceşti parteneri.
Distribuţia informaţiilor între bazele de date (localizate în Intranet) poate fi realizată prin intermediul Internet-ului [1] (figura 1). În acest caz, nu se mai poate presupune că sunt asigurate lărgime de bandă fără restricţii şi un înalt nivel de securitate; astfel, se impune utilizarea unor tehnologii care să permită eliminarea neajunsurilor oferite de modelele clasice (transfer mare de date, posibilitatea întreruperii legăturii etc.).

Tehnologia agenţilor mobili se impune în această situaţie datorită numeroaselor avantaje pe care le oferă [3]:
*1. Reducerea traficului de reţea.* Un agent mobil migrează printre serverele din sistem pentru a colecta informaţiile dorite de utilizator şi pentru a le aduce pe serverul de pe care a plecat. În acest mod, traficul datorat transmiterii datelor prin reţea este redus, deoarece agenţii procesează local volumul mare de date şi transportă doar datele esenţiale care să fie prezentate utilizatorului.
*2. Flexibilitatea.* Agenţii mobili sunt procese flexibile, care pot fi configurate la lansarea în execuţie, eliminând astfel necesitatea de a actualiza soft-urile de la nivelul fiecărei firme. De exemplu, agenţii pot fi proiectaţi să colecteze doar anumite tipuri de informaţii necesare unui anumit utilizator, sau să selecteze următorul server pe care îl vizitează în funcţie de datele colectate de la server-ul curent.
*3. Reducerea timpului de căutare (obţinere a informaţiilor).* În general, este necesară trimiterea mai multor cereri pentru a obţine datele dorite din multitudinea de informaţii disponibile la nivelul unei firme. Deoarece pentru a obţine rezultatele dorite agenţii filtrează datele primite în urma interogării unui server, prin reţea sunt transmise doar datele necesare de la nivelul unei firme. Astfel, agenţii mobili sunt mai eficienţi decât schimbul de mesaje pentru obţinerea informaţiilor despre un produs de la mai multe firme.
*4. Siguranţa.* Agenţii mobili pot acţiona şi în cazul legăturilor Internet intermitente.

Utilizarea agenţilor mobili în comerţul electronic este un subiect de cercetare foarte des abordat în ultima perioadă. Cercetările în care agenţii sunt utilizaţi în B2B au în vedere următoarele 5 direcţii de bază: descoperirea serviciilor, medierea, negocierea, managementul proceselor (fie flux de lucru sau lanţ de distribuţie) şi evaluarea.
În cazul sistemului B2B_RFID se propune utilizarea agenţilor mobili pentru:

1. implementarea funcţiei de căutare a informaţiilor despre un produs, începând cu momentul înregistrării lui în sistem până în momentul curent (determinarea trasabilităţii produsului);
2. implementarea opţiunii de căutare a unui produs sau serviciu care îndeplineşte caracteristicile specificate de un utilizator la un moment dat.

**Arhitectura sistemului multiagent**

Pentru a integra tehnologia agent în cadrul sistemului existent, trebuie respectate două cerinţe[4]:

- ➢ la nivelul fiecărei firme să ruleze o platformă de agenţi, care să reprezinte mediul de execuţie pentru agenţii din cadrul sistemului;
- ➢ la nivelul fiecărei platforme să ruleze în permanenţă un agent supervizor, care să coordoneze activitatea celorlalţi agenţi.

Arhitectura sistemului multiagent prezentată în figura 2, cuprinde atât agenţi mobili, cât şi agenţi statici [5]. Agenţii statici oferă resurse şi facilitează accesul la acestea pentru agenţii mobili. Agenţii mobili migrează intre diferitele servere din sistem şi utilizează aceste resurse pentru a-şi îndeplini scopurile. Având în vedere problemele care le ridică migrarea agentului de la o firmă la alta, în cadrul sistemului trebuie asigurat şi un grad minim de securitate şi siguranţă.
În continuare se prezintă agenţii consideraţi şi rolurile lor în cadrul sistemului multiagent propus.

*1. Agentul Supervizor* este un agent static care coordonează evenimentele generate la nivelul unei firme. Serviciile oferite de acesta au în vedere resursele de informaţii, utilizatorii şi agenţii de pe platforma de la nivelul firmei. Agenţii supervizori sunt responsabili pentru deplasarea către o firma destinaţie şi asigurarea transferului cu succes a agentului mobil. De asemenea, el autentifică şi efectuează o verificare de validare pentru agenţii mobili care doresc să ruleze pe platforma de la nivelul firmei. Agenţii care nu se pot autentifica sau nu trec de testul de validare sunt respinşi. Agentul mobil sosit pe platforma curentă va fi lansat în execuţie de către agentul supervizor într-un mediu de execuţie corespunzător nivelului de încredere care i-a fost asociat la configurarea lui. Agenţii de la firmele care prezintă încredere pot accesa mai multe informaţii decât agenţii de la firmele necunoscute. Mediul de execuţie potrivit va depinde de nivelul de acces acordat agentului mobil şi de funcţiile pe care acesta doreşte să le execute.
Pe lângă asigurarea securităţii, agentul supervizor are şi funcţia de manager central în cadrul unei platforme şi este responsabil cu ţinerea evidenţei agenţilor care se execută la un moment dat, comunicaţia între agenţi şi resursele de informaţii, precum şi comunicaţia între agenţii existenţi la nivelul platformei.

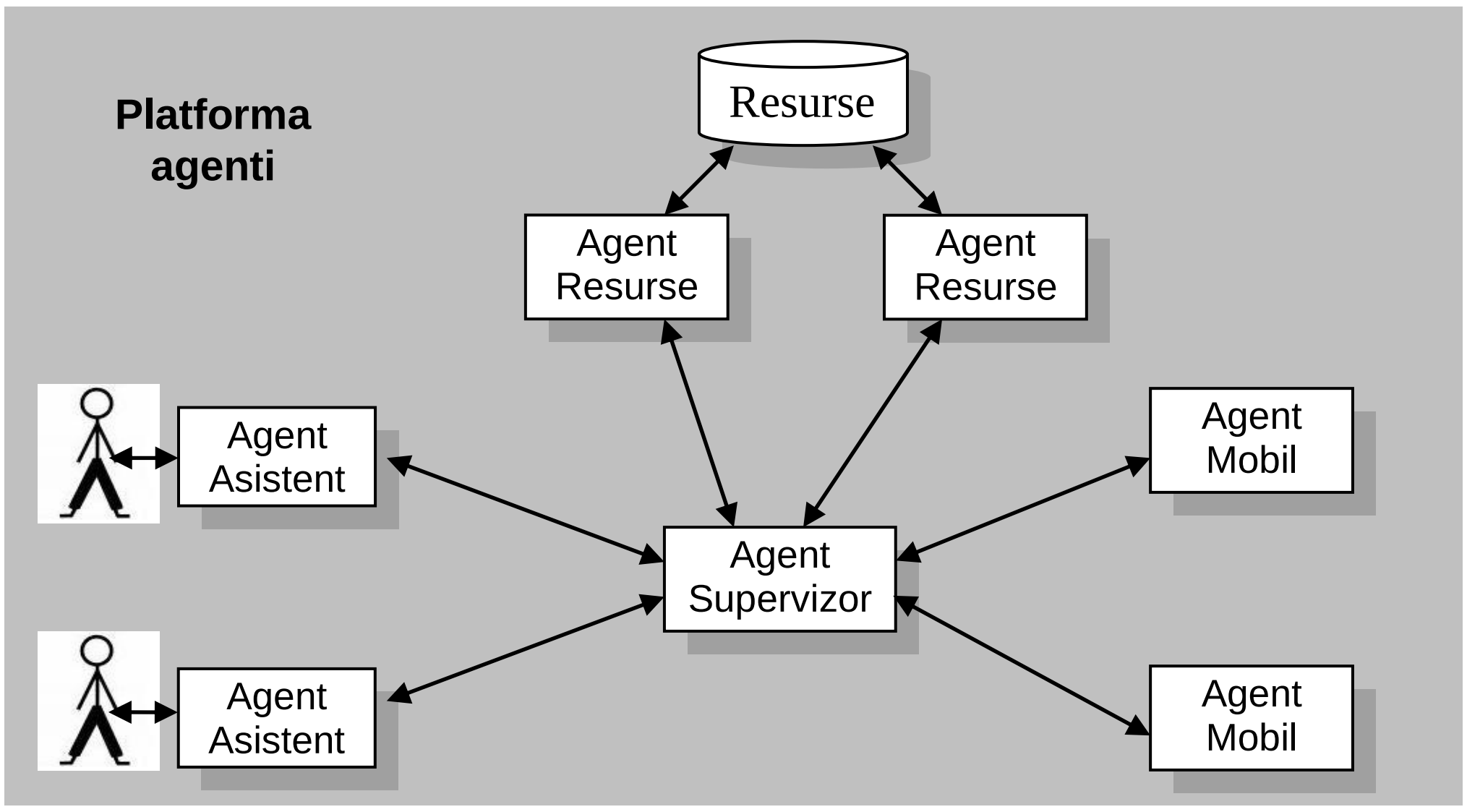

**Figura 2. Arhitectura sistemului multiagent de la nivelul unei firme**

2. *Agenţii de resurse* sunt agenţi statici care furnizează un nivel de abstractizare între o resursă de informaţie şi agentul mobil solicitant. Scopul lui este de a media accesul unui agent mobil la o resursă particulară de informaţie, deoarece ştie cum să o acceseze şi, de asemenea, „înţelege” structura de permisiuni asociate ei.
Interacţiunea dintre agenţii de resurse şi agenţii mobili formează punctul central al acestei arhitecturi. Flexibilitatea pe care o au agenţii mobili în accesul la resursele de informaţii şi modul în care interpretează rezultatele vor determina utilitatea lor. Agenţii de resurse sunt dezvoltaţi pentru a interfaţa bazele de date.

3. *Agenţii mobili* sunt componentele din cadrul arhitecturii care pot migra între platformele asociate firmelor. Ei reprezintă mecanismul prin care utilizatorii obţin accesul la resursele de informaţii dorite.
Iniţial, un utilizator lansează un agent mobil în cadrul unei anumite platforme, numită platformă gazdă. Agenţii mobili sunt echipaţi cu un set de scopuri specifice task-ului utilizatorului, care descriu natura şi limitele acţiunilor lor. Pe lângă limitele în funcţionalitate pe care utilizatorii le impun asupra lor, este foarte probabil ca ei să întâlnească noi limite care există în cadrul platformelor vizitate (sub forma normelor de securitate, autentificare, validare sau alte restricţii). În unele cazuri, aceste limite vor împiedica îndeplinirea scopurile atribuite acestuia. În momentul actual încă nu s-a găsit o metodă viabilă care să asigure faptul că un agent nu este compromis pe parcursul traseului sau că un agent mobil nu va avea un comportament diferit de cel primit la lansare. Prin urmare, ei trebuie să fie monitorizaţi în permanenţă de agentul supervizor şi de agenţii de resurse şi să primească cât mai puţine permisiuni. Un sistem cuprinzător de autentificare, validare şi permisiuni de acces este de o foarte mare importanţă pentru a asigura faptul ca agenţii rău intenţionaţi nu pot aduce daune sistemului, iar agenţii legitimi pot să îşi îndeplinească scopurile [6].
4. *Agentul de interfaţă* oferă un nivel de abstractizare utilizatorului, care nu trebuie să fie conştient de existenta arhitecturii de agenţi. Principalele sarcini ale acestui agent sunt:

- lansarea în execuţia a agenţilor mobili ca răspuns a unei cereri din partea utilizatorului;
- organizarea şi preprocesarea informaţiilor colectate de agentul mobil într-o formă corespunzătoare pentru utilizator. Acest lucru poate implica filtrarea informaţiilor replicate, prezentarea informaţiilor urgente imediat ce sunt disponibile, redarea informaţiilor utilizând instrumente cu care utilizatorul este obişnuit.

Interfaţa agentului mobil va fi realizată cu un grad ridicat de generalitate, astfel încât să permită utilizatorului să introducă criterii specifice pentru căutare, precum şi modul de căutare. De exemplu, printre condiţiile de căutare se pot considera: tipuri de produse, furnizorii de un anumit produs, data de fabricaţie etc. La modul de căutare se vor putea specifica: serverele care sa fie vizitate etc. (de exemplu, să fie vizitate serverele care deţin produse furnizate de un anumit producător).

Într-o arhitectură generală se stabileşte un itinerar pentru agentul mobil, considerându-se algoritmi de determinare a drumului optim, indicându-se ordinea de vizitare a site-urilor B2B. Un caz particular al acestei arhitecturi este cel în care se doreşte determinarea trasabilităţii produselor. In acest caz itinerarul este deja stabilit şi nu se poate modifica. Pornind de la destinaţia unui produs se vizitează pe rând toate firmele care l-au avut în stoc la un moment dat, ordinea fiind determinată în mod unic.

Pentru a alege platforma de agenţi care să fie utilizată pentru implementarea agenţilor au fost luate în considerare iniţial câteva criterii eliminatorii pentru a simplifica procesul de evaluare şi pentru a direcţiona eforturile către platformele agent cu posibilităţi mai mari pentru dezvoltări ulterioare. Astfel, au fost eliminate de la început platformele care nu îndeplinesc una din următoarele condiţii [7]:

- să ofere suport atât pentru sistemul de operare Windows, cât şi pentru Unix;
- să respecte standardele impuse în domeniu;
- să ofere o documentaţie minimală;
- să ofere suport pentru agenţi mobili;
- să fie disponibilă o versiune de testare (full functional, limited, student).

Având în vedere criteriile enumerate mai sus, au fost selectate următoarele platforme care le îndeplinesc pe toate:

*Grasshopper*

- respectă ambele standarde specifice sistemelor multi-agent (FIPA şi MASIF);
- are posibilitatea de a simula mobilitatea de tip strong (o caracteristică foarte importantă pentru a realiza aplicaţii în care agentul migrează, iar platforma care l-a creat poate fi deconectată de la reţea);
- ca dezavantaj, nu a mai fost dezvoltat din noiembrie 2003.

*Aglets*

- este utilizat în multe proiecte;
- este special pentru agenţi mobili;
- nu are dezvoltată o versiune care să ruleze pe dispozitive mobile.

*Voyager*

- timpul necesar trimiterii unui mesaj este foarte mic, ceea ce este foarte important, deoarece într-o aplicaţie multi-agent, schimbul de mesaje este foarte frecvent;
- oferă suport pentru CORBA şi poate fi utilizat în aplicaţii de integrare a sistemelor vechi;
- nu are dezvoltată o versiune care să ruleze pe dispozitive mobile.

*JADE*

- este o platformă gratis;
- este utilizată în foarte multe proiecte care necesită platforme multi-agent;
- este în continuă dezvoltare;
- oferă versiunea LEAP pentru dispozitivele mobile cu resurse limitate de calcul.

Din specificaţiile prezentate mai sus se observă ca platforma JADE este cea mai potrivită pentru necesităţile proiectului. Un criteriu esenţial în alegerea ei a fost existenţa unei versiuni care se poate executa pe dispozitive mobile (PDA, telefon mobil). În acest mod, accesul la funcţiile elementare ale sistemului nu mai este dependent de disponibilitatea unui calculator.

## Concluzii

În cadrul acestei lucrări a fost prezentată arhitectura unui sistem multiagent propus a fi integrat într-un sistem B2B, pentru a permite extragerea cunoştinţelor. Acest sistem va fi implementat utilizând agenţi mobili datorită numeroaselor avantaje oferite în comparaţie cu alte tehnologii. Sistemul considerat se adresează provocărilor din domeniul B2B lansate de apariţia noilor cerinţe, de volumul exploziv de informaţii şi de problemele legate de performanţa aplicaţiilor. Astfel, în cadrul sistemului B2B_RFID, sistemul multiagent va fi utilizat concret pentru determinarea trasabilităţii unui produs şi pentru căutarea entităţilor care respectă anumite criterii specificate de un utilizator.

## Referinţe bibliografice